\documentstyle[12pt]{article}
\begin{document}
\def\theequation{\arabic{equation}}
\newcommand{\be}{\begin{equation}}
\newcommand{\ee}{\end{equation}}
\begin{titlepage}
\title{\bf Scalar gravitational waves \\
and Einstein frame}
\author{S. Bellucci$^1$, V. Faraoni$^{1,2}$ and D. Babusci$^1$\\ \\
{\small \it $^1$ INFN-Laboratori Nazionali di Frascati, 
P.O. Box 13, I-00044 Frascati, Roma (Italy)}\\
{\small \it $^2$ Physics Department, University of Northern British Columbia}\\
{\small \it 3333 University Way, Prince George, BC V2N~4Z9 (Canada)}
}
\date{}
\maketitle
\thispagestyle{empty}      \vspace*{1truecm}
\begin{abstract}
The response of a gravitational wave detector to scalar waves is
analysed in the framework of the
debate on the choice of conformal frames for scalar-tensor theories. A 
correction to the geodesic equation arising in the Einstein
conformal frame modifies the geodesic deviation equation. This
modification is due to the non-metricity of the theory in the Einstein
frame, yielding a longitudinal mode that is absent in the Jordan
conformal frame.
\end{abstract} \vspace*{0.3truecm} 
\begin{center}
\end{center}

\end{titlepage}  

setcounter{equation}{0}

Virtually every modern theoretical attempt to unify gravity with the
remaining interactions requires the introduction of scalar fields
(e.g. the dilaton and the moduli fields in string theory 
\cite{GSW}).

In the literature on scalar fields it is claimed that the Jordan frame version of Brans-Dicke
and scalar-tensor theories is untenable, owing to the problem
of negative kinetic energies \cite{OR,VFIJTP}.
In turn, the Einstein frame version of
scalar-tensor theories $-$ obtained by a conformal rescaling
$g_{\mu\nu}\to{\tilde g}_{\mu\nu}$ and a nonlinear field redefinition
$\phi\to{\tilde{\phi}}$ $-$ has a positive definite energy.
In this frame however, there is
a violation of the Equivalence Principle, due to an anomalous coupling
of the scalar field $\phi$ to ordinary matter. Naturally, this
violation is small and compatible with the available tests of the
Equivalence Principle.\footnote{
For the consequences of such tests in a more specific framework,
see e.g. \cite{new1}.}
It is, indeed, a low-energy manifestation
of compactified theories \cite{Cho,TV,DP}.

Acknowledging the debate in the literature about the conformal frames
used for the description of scalar-tensor theories, we focus on
possible phenomenologically distinctive features emerging in the
conformal frames. As an example of such phenomena, we consider in
this Letter the interaction of scalar waves with a gravitational wave
detector, as seen in the Einstein frame. The same 
question has been addressed in the Jordan frame \cite{MaggNic}.

If the scalar fields are required for unifying the fundamental
interactions among particles at high energies, then they
must be present in the early universe. Although a fundamental
cosmological scalar field may have settled to a constant value  during the
era of matter domination \cite{DamourNordvedt}, or even during inflation
\cite{Garcia}, it would leave an imprint  in the cosmological background
of gravitational waves by contributing spin zero modes. 

At this point, the question arises, whether the gravitational wave 
detectors presently existing (i.e. the resonant detectors) or under
construction (i.e. the interferometric detectors) can detect relic 
(cosmological) scalar gravitational waves\footnote{
For a recent review of the status of gravitational wave detectors, see
\cite{GWD}; see also \cite {new2}.}.
As we will see, a correction to the 
geodesic equation arises in the Einstein frame and modifies the 
geodesic deviation equation.

The main motivation for this analysis arises from cosmology: given the
unavoidability of scalar fields in the early universe, scalar modes must
be present in the relic gravitational wave background of cosmological
origin, together with spin two modes. For simplicity, the prototype of
scalar-tensor theories, i.e. Brans-Dicke \cite{BD} theory, is used
in the following; however, the analysis below
extends immediately to generalized scalar-tensor gravity.

We plan our paper as follows. We begin by briefly discussing
the essential features of the two conformal frames. Next,
we compute the corrections to the geodesic and
geodesic deviation equations that appear in the Einstein frame; 
finally, we conclude with a discussion of the results presented
here and the comparison of our approach with the string formulation.

In the usual formulation of Brans-Dicke theory \cite{BD} in the so-called
Jordan frame, gravitational waves are represented by the metric and
scalar fields\footnote{We adopt the notations and conventions of
\cite{Wald}.}
\be  \label{Jgw}
g_{\mu\nu} = \eta_{\mu\nu} + h_{\mu\nu} \, , \,\,\,\,\,\;\;\;\;\;\;\;
\phi=\phi_0 + \varphi \; ,
\ee
where ${\cal O}(h_{\mu\nu}) = {\cal O}(\varphi/\phi_0) = {\cal O}(\epsilon)$, 
$\epsilon$ being a smallness parameter. The linearized field equations in 
a region outside sources are \cite{Will}
\be  \label{Jlinearized1}
R_{\mu\nu} = T_{\mu\nu}\left[ \varphi \right] +
{\cal O}( \epsilon^2 ) = \frac{\partial_{\mu} \partial_{\nu}
\varphi}{\phi_0} + {\cal O}( \epsilon^2 ) \; ,
\ee
\be  \label{Jlinearized2}
\Box \varphi = 0 \; .
\ee
The energy density of the scalar waves seen by an observer with
four-velocity $u^{\mu}$ is 
$\rho \equiv T_{\mu\nu} [\varphi] u^{\mu} u^{\nu}$ and its sign 
is indefinite: for example, for a monochromatic scalar wave 
$\varphi = \varphi_0 \cos (k_{\alpha} x^{\alpha})$, one has 
$\rho = - (k_{\alpha} u^{\alpha})^2 \varphi/\phi_0$. 

Brans-Dicke theory can be reformulated in the Einstein conformal frame
\cite{BD} by means of the conformal transformation
\be  \label{conformaltransformation}
g_{\mu\nu} \longrightarrow  \tilde{g}_{\mu\nu}=\Omega^2 g_{\mu\nu}\; ,
\;\;\;\;\;\; \Omega=\sqrt{G \phi} \; , 
\ee
and of the scalar field redefinition
\be  \label{Escalar}
\tilde{\phi} = \frac{1}{\chi}\,\ln \left(\frac{\phi}{\phi_0}\right)\;, 
\qquad  \chi \equiv \left(\frac{16\pi G}{2\omega + 3}\right)^{1/2}
\ee
where $\omega$ is the Brans-Dicke parameter.
Scalar-tensor gravitational waves are described in the Einstein frame by
\be  \label{Egw}
\tilde{g}_{\mu\nu}=\eta_{\mu\nu} + \tilde{h}_{\mu\nu} \, ,
\,\,\,\,\,\;\;\;\;\;\;\; \tilde{h}_{\mu\nu}=h_{\mu\nu}
+\frac{\varphi}{\phi_0} \, \eta_{\mu\nu} \; , \;\;\;\;\;\;\;\;
\tilde{\phi}=\tilde{\phi}_0+\tilde{\varphi} \; ,
\ee
where one can use eq.~(\ref{Escalar}) to express
the field $\tilde{\varphi}$ through $\phi$
\be
\tilde{\varphi}=\frac{1}{\chi}\,\frac{\varphi}{\phi_0}  \; .
\ee
The linearized field equations outside sources are\footnote{
Note that the extension of our treatment and results to
the most general scalar-tensor case is straightforward at this point,
as the necessary modifications in the field equations only affect
terms of higher order in the fields. Hence, within the linearized
approximation adopted here, the non-geodesic term found below
(eq.~(\ref{Egeodesicdeviation})) is the
same for all scalar-tensor theories.}
\be  \label{Elinearized1}
\tilde{R}_{\mu\nu}-\frac{1}{2} \tilde{g}_{\mu\nu}
\tilde{R}=8\pi \left( \tilde{T}_{\mu\nu} \left[ \tilde{\varphi} \right] +
T^{(eff)}_{\mu\nu} \left[ \tilde{h}_{\alpha\beta} \right] \right) \; ,
\ee
\be  \label{Elinearized2}
\Box \tilde{\varphi} = 0 \; ,
\ee
where
\be  \label{canonical}
\tilde{T}_{\mu\nu} \left[ \tilde{\varphi} \right] =
\partial_{\mu} \tilde{\varphi} \, \partial_{\nu}
\tilde{\varphi} -\frac{\eta_{\mu\nu}}{2} \,  
\partial^{\alpha}\tilde{\varphi}\, 
 \partial_{\alpha} \tilde{\varphi}
\ee
and $T^{(eff)}_{\mu\nu} [\tilde{h}_{\alpha\beta}] $ is
Isaacson's effective stress-energy tensor \cite{MTW}
(it yields the contribution of the tensor modes
$\tilde{h}_{\alpha\beta}$ and is only of order $\epsilon^2$). In the
Einstein frame both scalar and tensor waves
yield second order contributions to the effective energy density and the
canonical form (\ref{canonical}) of 
$\tilde{T}_{\mu\nu} \left[ \tilde{\varphi} \right] $ (as opposed to 
$ T_{\mu\nu} \left[ \varphi \right] $ in eq.~(\ref{Jlinearized1})) shows
that the scalar contribution
is non-negative (for a monochromatic plane wave
$\tilde{\varphi} = \tilde{\varphi}_0 \cos (l^{\alpha} x_{\alpha})$ 
one has $\tilde{\rho}_{\tilde{\varphi}} =
\tilde{T}_{\mu\nu} u^{\mu} u^{\nu} = (l_{\alpha} u^{\alpha} \varphi)^2 
\geq 0 $).\footnote{We do not
address here the question of whether the reformulated theory is the
physical one, as opposed to its Jordan frame counterpart.
Our purpose is to obtain some phenomenological
consequence of the Einstein reformulation of the
theory, for the spectrum of scalar gravitational waves.}
For this reason, the Einstein frame is often used to compute the energy
density of the stochastic background of scalar waves (e.g. \cite{MaggNic});
this procedure implicitly assumes that the Jordan and the Einstein frames
are physically equivalent. If it was true, this equivalence would imply
that the physics is the same in both frames, while we will see later
that this is not the case for the geodesic deviation equation,
which provides the theoretical ground for describing the
response of a gravitational wave detector to a scalar wave.

Next, we set out to determine the corrections to the geodesic and 
geodesic deviation equations due to the reformulation of the 
theory to the Einstein frame. A convenient starting point is the 
conservation law for the matter energy-momentum tensor 
$\tilde{T}_{\mu\nu}^{(m)}$ in the Einstein frame \cite{Wald}
\be  \label{conservation}
\tilde{\nabla}_{\mu} \tilde{T}^{(m) \, \mu\nu}=-\frac{1}{\Omega}
\frac{\partial \Omega}{\partial \phi} \tilde{T}\, \tilde{\nabla}^{\mu}
\phi = -\frac12\,\chi\,\tilde{T}\,\tilde{\nabla}^{\mu}\tilde{\phi} \;,
\ee
where $\tilde{T} = {\tilde{T}^{(m)}}{}_{\mu}{}^{\mu}$. For a dust fluid, 
i.e. for $T_{\mu\nu}^{(m)}= \rho u_{\mu} u_{\nu}$ 
(with $u_{\alpha}u^{\alpha} = -1$), one obtains the correction to the 
geodesic equation for a massive test particle
\be  \label{Egeodesic}
\frac{Dp^{\mu}}{D\lambda}=\frac{d^2 x^{\mu}}{d\lambda^2}+
\tilde{\Gamma}^{\mu}_{\alpha\beta} \frac{dx^{\alpha}}{d\lambda} 
\frac{dx^{\beta}}{d\lambda} =\frac12\,\chi\,\partial^{\mu} \tilde{\phi} \; ,
\ee
where $p^{\mu}\equiv dx^{\mu}/d\lambda$ is the four-tangent to the world
line of  a test particle.  The correction is a force that couples
universally to massive test particles (e.g. \cite{Cho}); the equation of
null geodesics instead is conformally invariant\footnote{The equation of
null geodesics is obtained as the high frequency limit of the Maxwell
equations, which are conformally invariant in four spacetime dimensions.
Alternatively, for the Maxwell field one has $\tilde{T}=0$ in
eq.~(\ref{conservation}). In a different context, superconformal
invariance was proposed to explain why newtonian gravity shadows
cosmological constant effects \cite{new3}.}  
and is the same in the Jordan and in the Einstein frame.

A term with a
formal similarity to our correction of the geodesic equation has been
obtained also in a model inspired by string theory
\cite{Gasperini99}. However, while the string dilaton in general couples
differently to bodies with different internal nuclear structure, which
carry a dilatonic charge \cite{Gasperini99}, the Brans-Dicke field of  
scalar-tensor theories couples in the same way to every form of matter
which has an energy-momentum tensor with a nonvanishing trace (cf.
eq.~(\ref{conservation})). In scalar-tensor theories there is no need to
make assumptions on the form of the different couplings and on dilatonic
charges, due to the universality of the coupling. See below for
a comment on the distinct physical implications of our scalar-tensor
case, with respect to the string inspired model of Ref. [15]

One can now derive the correction to the geodesic deviation equation; let
$\left\{ \gamma_s ( \lambda ) \right\}$ be a smooth one-parameter family
of test particle worldlines ($\lambda$ parameterizes the position along the
worldline and $s$ identifies curves in the family). If $p^{\alpha}=\left(
\partial /\partial \lambda \right)^{\alpha} $, $ s^{\beta}=
\left( \partial /\partial s \right)^{\beta} $, one has $ \partial
p^{\alpha}/\partial s =\partial s^{\alpha} / \partial \lambda $ and 
$ D p^{\alpha}/ D s =D s^{\alpha} /D \lambda $, where $D/Ds \equiv
s^{\alpha}\nabla_{\alpha} $, $ D/D\lambda \equiv p^{\beta}\nabla_{\beta}
$. The relative acceleration of two neighbouring curves in $\left\{
\gamma_s 
\right\}$ is found to be
\be  \label{Egeodesicdeviation}
\tilde{a}^{\alpha}= \tilde{R}_{\beta\delta\gamma}{}^{\alpha} s^{\beta}
p^{\gamma}
p^{\delta} + \frac12\,\chi\,\frac{D}{Ds}
\left( \partial^{\alpha} \tilde{\phi} \right)
\ee 
(the calculation parallels the usual derivation of the geodesic deviation
equation in general relativity $-$ see e.g. \cite{Wald}). A similar
correction appears in string
theory, but it depends on the dilatonic charge \cite{Gasperini99}. Note
that the usual limit $\omega > 500 $, which would make the 
contribution of scalar waves in eq. (\ref{Egeodesicdeviation}) small, 
does not apply in the Einstein frame. In fact, such a limit on $\omega$ 
assumes that the Jordan frame formulation of Brans-Dicke theory is the 
relevant one, while we have abandoned it in favour of its Einstein frame 
counterpart.

From the geodesic deviation equation (\ref{Egeodesicdeviation}), we can 
calculate the time evolution of the separation $\Delta x^i$ between two 
neighboring test particles. In the proper frame of one of them  \cite{MTW} 
one has
\be 
\label{delta1}
\Delta \ddot{x^i} = {\tilde{R}_{j00}}{}^i\,\Delta x^j + \frac12\,\chi\,
\frac{\partial^2 \tilde{\varphi}}{\partial x_i \partial x_j}\,
\Delta x_j \;,
\ee
where the $t$-coordinate is the proper time of the particle at the frame's
origin, and we use the notation $\dot{w} \equiv \partial w/\partial t$. 
For a plane wave propagating along the $z$-axis, choosing the gauge 
$\theta_{0 \mu} = 0$, $\partial^{\mu} \theta_{\mu\nu} = 0$ (where 
$\theta_{\mu\nu} \equiv \tilde{h}_{\mu\nu}-1/2 \,\eta_{\mu\nu}
\, \tilde{h}^{\alpha}_{\alpha}+\tilde{\varphi} \,\eta_{\mu\nu} $), 
one obtains
\be  
\label{delta}
\Delta \ddot{x^i} = \frac12\,\left( \ddot{h}_{ij}  + \delta_{ij}\,\chi\, 
\ddot{\tilde{\varphi}} \right) \Delta x^j \;.
\ee
For a purely scalar gravitational wave, i.e. for $h_{ij} = 0$, 
the unit matrix in this equation implies the existence of three 
oscillations, two transversal (for $i=j=1,2$) and one longitudinal
(for $i=j=3$), in the scalar sector of the fluctuations of the metric, 
which are gravitationally coupled to the detector, through the 
geodesic deviation equation.

The transverse modes are already present
in the Jordan frame case, whereas the longitudinal oscillation is 
absent in the latter formulation of the theory (see \cite{MaggNic}).
The longitudinal mode arises due to the non-metricity of the
Brans-Dicke theory in the  Einstein frame, i.e. the fact that massive
particles do not follow geodesics of the metric $\tilde{g}_{\mu\nu}$.
Notice that this longitudinal mode is not a gauge artifact, as in Ref.
\cite{Shibataetal}, but is a physical effect. 
Moreover, Eq. (\ref{delta}) shows that the longitudinal mode of the
scalar waves
is as important as the transverse scalar modes $-$ the same coefficient
appearing as a common multiplicative factor for all scalar modes.

It is interesting also to examine more closely the relation of
our result with the string formulation \cite{Gasperini99}.
There is a formal correspondence between
the corrected geodesic deviation equation (\ref{delta1}),
and eq. (13) of Ref. \cite{Gasperini99}, after
the replacement $q \to \chi/2$ of the dilatonic charge $q$ with the quantity
$\chi/2$.
However, the choice $q$=0 is always possible in the model
of Ref.~\cite{Gasperini99}. This corresponds
to the only possibility to have, within that model, a universal
dilaton coupling. All other values of the dilaton coupling to matter
depend on the composition of the test particles. Hence, two different
gravitational wave detectors, made of distinct materials, would respond
differently to a scalar gravitational wave. Setting a vanishing
coupling $q$ in Ref. \cite{Gasperini99} yields no physical effect.
On the other hand, in our
case there is no way to get rid of the (composition independent)
physical effect by any choice of coupling, as $\chi$ cannot be set
to zero.

Irrespectively of whether the Einstein frame formulation is to
be preferred to its Jordan counterpart or not, one is motivated
at least by its occurrence in the literature in looking at
the effect of scalar gravitational waves in the Einstein version
of the theory. It is a fact $-$ and our result shows it clearly $-$
that in the latter there is a longitudinal effect associated with
a scalar wave, whereas such an effect is absent in the
Jordan frame formulation. From the phenomenological point of view,
we must account also for the necessary smallness of the violations of the 
Equivalence Principle in the Einstein frame formulation.

The main conclusion emerging from our work is that the Einstein frame
description of the interaction of a scalar gravitational wave with a
gravitational wave detector differs from the Jordan frame picture
\cite{MaggNic}, as made clear also from the string theory analysis of
Ref.~\cite{Gasperini99}. This fact testifies of the physical
inequivalence of the two conformal frames. 

\section*{Acknowledgments}

We wish to thank R. Jackiw for a useful remark on the definition
of the energy density in Einstein's theory of gravity.
SB acknowledges support by the ``B. Rossi" program of exchange MIT/INFN
and thanks warmly MIT-CTP for hospitality.


{\small \begin{thebibliography}{99}

\bibitem{GSW} M.B. Green, J. Schwarz and E. Witten, {\em Superstring
Theory} (Cambridge University Press, Cambridge, 1987).

\bibitem{OR} P. Teyssandier and P. Tourrenc, {\em J. Math. Phys.}
{\bf 24} 2793 (1983); L. Sokolowski, {\em Class. Quant. Grav.} {\bf 6}
59; 2045 (1989); D.I. Santiago  and A.S.  Silbergleit,
{\em Gen. Rel. Grav.} {\bf 32} 565 (2000).

\bibitem{VFIJTP} V. Faraoni and E. Gunzig, {\em Int. J. Theor. Phys.} 
{\bf 38} 217 (1999); V. Faraoni, E. Gunzig and P. Nardone, {\em Fund. Cosm.
Phys} {\bf 20} 121 (1999).

\bibitem{new1} S. Bellucci and V. Faraoni, {\em Phys. Rev. D}
{\bf 49} 2922 (1994);
{\em Phys. Lett. B} {\bf 377} 55 (1996);
S. Bellucci, {\em Proceedings of the International Europhysics Conference
on High Energy Physics}, (Jerusalem, 19-26 August 1997), edited by D.
Lellouch, G. Mikenberg, E. Rabinovici, (Springer, Berlin), page 918,
hep/ph-9710562.

\bibitem{Cho} Y.M. Cho, {\em Phys. Rev. Lett.} {\bf 68} 3133 (1992);
{\em Class. Quant. Grav.} {\bf 14} 2963 (1997).

\bibitem{TV} T.R. Taylor and G. Veneziano, {\em Phys. Lett. B} {\bf 213} 
450 (1988).

\bibitem{DP} T. Damour and A.M. Polyakov, {\em Nucl. Phys. B} {\bf 423} 
532 (1994).

\bibitem{MaggNic} M. Maggiore and A. Nicolis, {\em Phys. Rev. D} {\bf
62} 024004 (2000).

\bibitem{DamourNordvedt} T. Damour and K. Nordvedt, {\em Phys. Rev.
Lett.} {\bf 70} 2217 (1993); {\em Phys. Rev. D} {\bf 48} 3436 (1993).
 
\bibitem{Garcia} J. Garcia-Bellido and M. Quir\`os, {\em Phys. Lett. B} 
{\bf 243} 45 ( 1990); J.D. Barrow and K. Maeda,  {\em Nucl. Phys. B}  
{\bf 341} 294 (1990).

\bibitem{GWD} {\em Proceedings of the Third Edoardo Amaldi Conference},
S. Meshkov {\em et al.} eds. (Pasadena, 1999), in press.

\bibitem{new2} B. Caron {\em et al.}, {\em 1st International LISA
Symposium on Gravitational Waves}, (Oxfordshire, England, 9-12 Jul 1996),
{\em Class. Quant. Grav.} {\bf 14} 1461 (1997).

\bibitem{BD} C.H. Brans and R.H. Dicke, {\em Phys. Rev.} {\bf 124} 925 (1961).

\bibitem{Wald}  R.M. Wald, {\em General Relativity} (The University of
Chicago Press, Chicago, 1984).

\bibitem{Will} C.M. Will {\em Theory and Experiment in Gravitational
Physics} (Cambridge University Press, Cambridge, 1993).

\bibitem{MTW} C.M. Misner, K.S. Thorne and J.A. Wheeler, {\em
    Gravitation} (Cambridge University
Press, Cambridge, 1973).

\bibitem{new3} S. Bellucci, {\em Phys. Rev. D} {\bf 57} 1057 (1998);
{\em Fortschr. Phys.} {\bf 40} 393 (1992); {\em Prog. Teor. Phys.}
{\bf 78} 1176 (1987); see also S. Bellucci and D. O'Reilly,
{\em Nucl. Phys. B} {\bf 364} 495 (1991).

\bibitem{Gasperini99} M. Gasperini, {\em Phys. Lett. B} {\bf 470} 67 (1999).

\bibitem{Shibataetal} M. Shibata, K. Nakao and T. Nakamura, {\em Phys.
Rev. D} {\bf 50} 7304 ( 1994).

\end{thebibliography}}
\end{document}